\documentclass[12pt]{article}
\usepackage[utf8]{inputenc}
\usepackage[T1]{fontenc}   
\usepackage[hidelinks]{hyperref}       %
\usepackage{url}    
\usepackage{booktabs}
\usepackage{nicefrac}
\usepackage{microtype}
\usepackage{verbatim}
\usepackage{amsfonts}
\usepackage{amsmath}
\usepackage{esint}
\usepackage{graphicx}
\usepackage{times}
\usepackage{color}
\usepackage{multirow}
\usepackage{appendix}
\usepackage[authoryear]{natbib}
\usepackage{rotating}

\newcommand{\captionfonts}{\normalsize}

\makeatletter  
\long\def\@makecaption#1#2{%
  \vskip\abovecaptionskip
  \sbox\@tempboxa{{\captionfonts #1: #2}}%
  \ifdim \wd\@tempboxa >\hsize
    {\captionfonts #1: #2\par}
  \else
    \hbox to\hsize{\hfil\box\@tempboxa\hfil}%
  \fi
  \vskip\belowcaptionskip}
\makeatother   

\usepackage[acronym]{glossaries}
\newacronym{snn}{SNN}{spiking neural network}
\newacronym{rnn}{RNN}{recurrent neural network}
\newacronym{lif}{LIF}{leaky integrate-an-fire}
\newacronym{psp}{PSP}{postsynaptic potential}
\newacronym{stdp}{STDP}{spike timing dependent plasticity}
\newacronym{rtrl}{RTRL}{real-time recurrent learning} 
\newacronym{backprop}{Backprop}{back-propagation of error algorithm}

\begin{document}
\hspace{13.9cm}
\ \vspace{20mm}

\noindent {\LARGE SuperSpike: Supervised learning in multi-layer spiking neural networks}

\ \vspace{20mm}

\noindent {\bf \large Friedemann Zenke$^{\displaystyle 1, \displaystyle 2}$ \&
Surya Ganguli$^{\displaystyle 1}$} \\
{$^{\displaystyle 1}$Department of Applied Physics\\
  Stanford University\\
  Stanford, CA 94305\\
  United States of America\\
}\\
{$^{\displaystyle 2}$Centre for Neural Circuits and Behaviour\\
University of Oxford\\
Oxford\\
United Kingdom}\\
%


\thispagestyle{empty}
\begin{center} {\bf Abstract} \end{center}
	A vast majority of computation in the brain is performed by spiking neural
	networks. Despite the ubiquity of such spiking, we currently lack an
	understanding of how biological spiking neural circuits learn and compute
	{\it in-vivo}, as well as how we can instantiate such capabilities in
	artificial spiking circuits {\it in-silico}.  Here we revisit the problem of
	supervised learning in temporally coding multi-layer spiking neural
	networks.  First, by using a surrogate gradient approach, we derive
	SuperSpike, a nonlinear voltage-based three factor learning rule capable of
	training multi-layer networks of deterministic integrate-and-fire neurons to
	perform nonlinear computations on spatiotemporal spike patterns.  Second,
	inspired by recent results on feedback alignment, we compare the performance
	of our learning rule under different credit assignment strategies for
	propagating output errors to hidden units.  Specifically, we test uniform,
	symmetric and random feedback, finding that simpler tasks can be solved with
	any type of feedback, while more complex tasks require symmetric feedback.
	In summary, our results open the door to obtaining a better scientific
	understanding of learning and computation in spiking neural networks by
	advancing our ability to train them to solve nonlinear problems involving
	transformations between different spatiotemporal spike-time patterns.

\section{Introduction}

Neurons in biological circuits form intricate networks in which the primary mode of communication occurs through spikes.  The theoretical basis for how such networks are sculpted by experience to give rise to emergent computations remains poorly understood.  
Consequently, building meaningful spiking models of brain-like
neural networks in-silico is a largely unsolved problem.
In contrast, the field of deep learning has made remarkable progress in
building non-spiking convolutional networks which often achieve human-level performance at solving difficult tasks \citep{schmidhuber_deep_2015, lecun_deep_2015}. 
Even though the details of how these artificial rate-based networks are trained may arguably 
be different from how the brain learns, several studies have begun to draw
interesting parallels between the internal representations formed by deep neural
networks and the recorded activity from different brain regions
\citep{yamins_performance-optimized_2014, mcclure_representational_2016, mcintosh_deep_2016, marblestone_toward_2016}. 
A major impediment to deriving a similar comparison at the spiking level is that we currently lack efficient ways of training 
\glspl{snn}, thereby limiting their applications to mostly
small toy problems that do not fundamentally involve spatiotemporal spike time computations. 
For instance, only recently have some groups begun to train \glspl{snn} on
datasets such as MNIST \citep{diehl_unsupervised_2015,
guergiuev_biologically_2016, neftci_neuromorphic_2016, petrovici_pattern_2017}, 
whereas most previous studies have used smaller artificial datasets.

The difficulty in simulating and training \glspl{snn} originates from multiple
factors. 
First, time is an indispensable component of the functional form 
of a \gls{snn}, as even individual stimuli and their associated
outputs are spatiotemporal spike patterns, rather than simple spatial 
activation vectors. This fundamental difference necessitates the use of different cost functions 
from the ones commonly encountered in deep learning.
Second, most spiking neuron models are inherently non-differentiable at
spike time and the derivative of their output with respect to synaptic weights is zero at all
other times.
Third, the intrinsic self-memory of most spiking neurons introduced by the
spike reset is difficult to treat analytically.
Finally, credit assignment in hidden layers is problematic for two reasons:
(i)~it is technically challenging because 
efficient auto-differentiation tools are not available for most event-based
spiking neural network frameworks, and (ii)~the method of weight updates implemented by
the standard \gls{backprop} is thought to be biologically implausible
\citep{grossberg_competitive_1987, crick_recent_1989}.

Several studies of multi-layer networks which build on the notion of
``feedback alignment'' \citep{lillicrap_random_2016} have recently illustrated
that the strict requirements imposed on the feedback by backpropagation of error signals can be loosened
substantially without a large loss of performance on standard benchmarks like
MNIST \citep{lillicrap_random_2016, guergiuev_biologically_2016,
neftci_neuromorphic_2016, baldi_learning_2016, liao_importance_2015}.
While some of these studies have been performed using spiking networks, they
still use effectively a rate-based approach
in which a given input activity vector is interpreted as the firing rate
of a set of input neurons \citep{eliasmith_large-scale_2012, diehl_unsupervised_2015,
guergiuev_biologically_2016, neftci_neuromorphic_2016, mesnard_towards_2016}.
While this approach is appealing because it can often be related directly to
equivalent rate-based models with stationary neuronal transfer functions, it
also largely ignores the idea that individual spike timing may carry additional
information which could be crucial for efficient coding \citep{thalmeier_learning_2015, 
deneve_efficient_2016, abbott_building_2016, brendel_learning_2017} 
and fast computation \citep{thorpe_speed_1996, gollisch_rapid_2008}.

In this paper we develop a novel learning rule to train multi-layer \glspl{snn} of deterministic \gls{lif} neurons 
on tasks which fundamentally involve spatiotemporal spike pattern transformations.
In doing so we go beyond the purely spatial rate-based activation vectors prevalent in deep learning. 
We further study how biologically more plausible strategies for deep credit assignment across multiple layers generalize to the enhanced 
context of more complex spatiotemporal spike-pattern transformations.

\subsection{Prior work}
Supervised learning of precisely timed spikes in single neurons and networks without 
hidden units has been studied extensively. 
\citet{pfister_optimal_2006} have used a probabilistic
escape rate model to deal with the hard nonlinearity of the spike.
Similar probabilistic approaches have also been used to derive \gls{stdp}
from information maximizing principles \citep{bohte_reducing_2007, toyoizumi_spike-timing_2005}.
In contrast to that, ReSuMe \citep{ponulak_supervised_2009} and SPAN
\citep{mohemmed_span:_2012} are deterministic approaches which can
be seen as generalizations of the Widrow-Hoff rule to spiking neurons.
In a similar vein, the Chronotron \citep{florian_chronotron:_2012} 
learns precisely timed output spikes
by minimizing the Victor-Pupura distance \citep{victor_metric-space_1997}
to a given target output spike train. 
Similarly, \citet{gardner_supervised_2016}
and \citet{albers_learning_2016} have studied the convergence 
properties of rules that reduce the
van Rossum distance by gradient
descent. Moreover, \citet{memmesheimer_learning_2014} proposed 
a learning algorithm which achieves high capacity in learning
long precisely timed spike trains in single units and recurrent networks.
The problem of sequence learning in recurrent neural networks has 
also been studied as a variational learning problem
\citep{brea_matching_2013, jimenez_rezende_stochastic_2014} and 
by combining adaptive control theory with heterogeneous neurons
\citep{gilra_predicting_2017}.

Supervised learning in \glspl{snn} without hidden units has also been studied
for classification problems. For instance,
\citet{maass_real-time_2002} have used the p-delta rule
\citep{auer_learning_2008} to train the readout layer of a liquid state machine.
Moreover, the Tempotron \citep{gutig_tempotron:_2006, gutig_spiking_2016},
which can be derived as a gradient-based approach \citep{urbanczik_gradient_2009},
classifies large numbers of temporally coded spike patterns without explicitly specifying a target
firing time.

Only a few works have embarked upon the problem of training
\glspl{snn} with hidden units to process precisely timed input and output 
spike trains by porting \gls{backprop} to the spiking domain. 
The main analytical difficulty in these approaches arises from partial
derivatives of the form $\nicefrac{\partial S_i(t)}{\partial w_{ij}}$ where
$S_i(t)=\sum_k\delta(t-t^k_i)$ is the spike train of the hidden neuron $i$ and 
$w_{ij}$ is a hidden weight.
SpikeProp \citep{bohte_error-backpropagation_2002} sidesteps this problem by
defining a differentiable expression on the firing times instead, on which
standard gradient descent can be performed. 
While the original approach was limited to a single spike
per neuron, multiple extensions of the algorithm exist, some of which also
improve its convergence properties \citep{mckennoch_fast_2006,
booij_gradient_2005, shrestha_adaptive_2015, de_montigny_analytical_2016,
banerjee_learning_2016, shrestha_robust_2017}.
However, one caveat of such spike timing based methods is that they 
cannot learn starting from a quiescent state of no spiking, as the spike time is 
then ill-defined.
Some algorithms, however, do not suffer from this limitation. 
For instance, an extension of ReSuMe to multiple layers was proposed 
\citep{sporea_supervised_2013} in which error signals were backpropagated 
linearly. More recently, the same group proposed a more principled  
generalization of \gls{backprop} to \glspl{snn} in \citet{gardner_learning_2015} using
a stochastic approach, which can be seen as an
extension of \citet{pfister_optimal_2006} to multiple layers.
In a similar flavour as \citet{fremaux_functional_2010}, 
\citet{gardner_learning_2015} substitute the partial derivative of hidden spike trains by
a point estimate of their expectation value. 
Although, theoretically, stochastic approaches avoid problems arising from quiescent
neurons, convergence can be slow and the injected noise may
become a major impediment to learning in practice.
Instead of approximating partial derivatives of spike trains by their
expectation value, in \citet{bohte_error-backpropagation_2011} the corresponding partial 
derivative is approximated as a scaled Heaviside function of the membrane voltage.
However, due to the use of the Heaviside function, this approach has a vanishing
surrogate gradient for sub-threshold activations which limits the algorithm's
applicability to cases in which hidden units are not quiescent. 
Finally, \citet{huh_gradient_2017} proposed another interesting approach in
which, instead of approximating partial derivatives for a hard spiking nonlinearity, 
instead a ``soft'' spiking threshold is used, for which by design 
standard techniques of gradient descent are applicable.

In contrast to these previous works, our method permits to train multi-layer networks 
of \textit{deterministic} \gls{lif} neurons to solve tasks involving 
spatiotemporal spike pattern transformations without the need for injecting noise
even when hidden units are initially completely silent.
To achieve this, we approximate the partial derivative of the hidden unit outputs as
the product of the filtered presynaptic spike train and a nonlinear function of the
postsynaptic \textit{voltage} instead of the postsynaptic spike train. 
In the following section we explain the details of our approach.

\section{Derivation of the SuperSpike learning rule}

To begin, we consider a single \gls{lif}
neuron which we would like to emit a given target spike train $\hat{S}_{i}$ for
a given stimulus. Formally, we can frame this problem as an optimization
problem in which we want to minimize the van Rossum distance
\citep{van_rossum_novel_2001, gardner_supervised_2016} 
between $\hat{S}_{i}$ and the actual output spike train $S_{i}$, 
\begin{equation}
L=\frac{1}{2}\int_{-\infty}^{t}ds\left[\left(\alpha\ast\hat{S}_{i}-\alpha\ast S_{i}\right)(s)\right]^{2}
	\label{eq:cost}
\end{equation}
where $\alpha$ is a normalized smooth temporal convolution kernel. We
use double exponential causal kernels throughout because they can be easily
computed online and could be implemented as
electrical or chemical traces in neurobiology. 
When computing the gradient of Eq.~\ref{eq:cost}
with respect to the synaptic weights $w_{ij}$ we get 
\begin{equation}
\frac{\partial L}{\partial w_{ij}}=-\int_{-\infty}^{t}ds\left[\left(\alpha\ast\hat{S}_{i}-\alpha\ast S_{i}\right)(s)\right]\,\left(\alpha\ast\frac{\partial S_{i}}{\partial w_{ij}}\right)(s)\label{eq:gradient}
\end{equation}
in which the derivative of
a spike train $\nicefrac{\partial S_{i}}{\partial w_{ij}}$ appears. 
This derivative is problematic because for most neuron models it is zero except
at spike times at which it is not defined. 
Most existing training algorithms circumvent this problem by
either performing optimization directly on the membrane potential $U_{i}$ or by introducing noise
which renders the likelihood of the spike train
$\left\langle S_{i}(t)\right\rangle$ a smooth function
of the membrane potential. 
Here we combine the merits of both approaches by
replacing the spike train $S_{i}(t)$ with a continuous auxiliary function
$\sigma(U_i(t))$ of the membrane potential. 
For performance reasons, we choose $\sigma(U)$ to be the negative side of a fast sigmoid 
(Methods), but other monotonic functions which increase steeply and peak
at the spiking threshold (e.g.\ exponential) should work as well. 
Our auxiliary function yields the replacement 
\begin{equation}
	\frac{\partial S_{i}}{\partial w_{ij}} \quad \rightarrow \quad \sigma^{\prime}(U_{i})\frac{\partial U_{i}}{\partial w_{ij}}.
\end{equation}
To further compute the derivative $\nicefrac{\partial U_{i}}{\partial w_{ij}}$
in the expression above, we exploit the fact that for current-based \gls{lif} models
the membrane potential $U_{i}(t)$ can be written in integral form
as a spike response model (SRM$_0$ \citep{gerstner_neuronal_2014}):
\begin{equation}
U_{i}(t)=\sum_{j}w_{ij}\left(\epsilon\ast S_{j}(t)\right)+\left(\eta\ast S_{i}(t)\right),\label{eq:srm}
\end{equation}
where we have introduced the causal membrane kernel
$\epsilon$ which corresponds to the \gls{psp}~shape and $\eta$ which captures
spike dynamics and reset. Due to the latter, $U_{i}$ depends on its own past
through its output spike train $S_{i}$. While this dependence does not allow us
to compute the derivative $\frac{\partial U_{i}}{\partial w_{ij}}$ directly, it
constitutes only a small correction to $U_i$ provided the firing rates are low.
Such low firing rates not only seem physiologically plausible, but also can be
easily achieved in practice by adding homeostatic mechanisms that
regularize neuronal activity levels. Neglecting the second term
simply yields the filtered presynaptic activity
$\frac{\partial U_{i}}{\partial w_{ij}}\approx(\epsilon\ast
S_{j}(t))$ which can be interpreted as the concentration of neurotransmitters at the synapse. Substituting this approximation back into Eq.~\ref{eq:gradient}, the gradient descent learning rule for a single neuron takes the form
\begin{equation}
\frac{\partial w_{ij}}{\partial t}=r\,\int_{-\infty}^{t}ds\,\underbrace{e_{i}(s)}_{\text{Error signal}}\,\underbrace{\alpha\ast\left(\underbrace{\sigma^{\prime}(U_{i}(s))}_{\text{Post}}\,\underbrace{\left(\epsilon\ast S_{j}\right)(s)}_{\text{Pre}}\right)}_{\equiv\lambda_{ij}(s)},
	\label{eq:learning_rule}
\end{equation}
where we have introduced the learning rate~$r$ and short notation for the output error signal $e_{i}(s)\equiv\alpha\ast(\hat{S}_{i}-S_{i})$ and
the eligibility trace $\lambda_{ij}$. In practice we evaluate the expression on minibatches
and we often use a per-parameter learning rate $r_{ij}$ closely related to
RMSprop \citep{hinton_neural_2012} to speed up learning.

Equation~\ref{eq:learning_rule} corresponds to the SuperSpike learning rule for
output neuron $i$. However, by redefining the error signal $e_i$ as a feedback
signal, we will use the same rule for hidden units as well.
Before we move on to testing this learning rule, we first 
state a few of its noteworthy properties: 
(i)~it has a Hebbian term which combines pre- and postsynaptic activity in a
multiplicative manner, 
(ii)~the learning rule is voltage-based,
(iii)~it is a nonlinear Hebbian rule due to the occurrence of $\sigma^{\prime}(U_{i})$, 
(iv)~the causal convolution with $\alpha$ acts as an eligibility trace to solve
the distal reward problem due to error signals arriving \emph{after} an error
was made \citep{izhikevich_solving_2007}, 
and, (v) it is a three factor rule in which the error signal plays the role
of a third factor \citep{fremaux_neuromodulated_2016, kusmierz_learning_2017}. Unlike most existing
three-factor rules, however, the error signal is specific to the postsynaptic
neuron, an important point which we will return to later.

\section{Methods}
We trained networks of spiking \gls{lif} neurons using a supervised
learning approach which we call ``SuperSpike''. This approach generalizes
the back propagation of error algorithm \citep{schmidhuber_deep_2015} 
as known from the multi-layer perceptron to deterministic spiking neurons.
Because the partial derivative and thus the gradient of deterministic spiking
neurons is zero almost everywhere, to make this optimization problem solvable,
we introduce a non-vanishing surrogate gradient \citep{hinton_neural_2012,
bengio_estimating_2013} (cf.~Eq.~\ref{eq:learning_rule}).
All simulations were run with a temporal resolution of 0.1ms using the Auryn
simulation library which is publicly available \citep{zenke_limits_2014}.

\subsection{Neuron model}
\label{sub:neuron_model}
We use \gls{lif} neurons with current-based synaptic
input because they can be alternatively formulated via their integral
form (cf.\ Eq.~\ref{eq:srm}). However, to simulate the membrane dynamics
we computed the voltage $U_{i}$ of neuron $i$ as described by 
the following differential equation
\begin{equation}
\tau^{{\rm mem}}\frac{dU_{i}}{dt}=(U^{\mathrm{rest}}-U_{i})+I_{i}^{\mathrm{syn}}(t)\label{eq:mem}
\end{equation}
in which the synaptic input current $I_{i}^{\mathrm{syn}}(t)$ evolves
according to
\begin{eqnarray}
\frac{d}{dt}I_{i}^{\mathrm{syn}}(t) & = & -\frac{I_{i}^{\mathrm{syn}}(t)}{\tau^{{\rm syn}}}+\sum_{j\in\mathrm{pre}}w_{ij}S_{j}(t) \quad .
\label{eq:synaptic_input}
\end{eqnarray}
The value of $I_{i}^{\mathrm{syn}}(t)$ jumps by an amount $w_{ij}$
at the moment of spike arrival from presynaptic neurons $S_{j}(t)=\sum_{k}\delta(t-t_{j}^{k})$
where $\delta$ denotes the Dirac $\delta$-function and $t_{j}^{k}$
($k=1,2,\cdots$) are firing times of neuron $j$. An action potential
is triggered when the membrane voltage of neuron $i$ rises above
the threshold value $\vartheta$ (see Table~\ref{tab:neuron-model-parameters}
for parameters). Following a spike the voltage $U_{i}$ remains clamped at
$U_{i}^{\mathrm{rest}}$ for $\tau^\mathrm{ref}=5\mathrm{ms}$ to emulate a refractory period.
After generation, spikes are propagated to other neurons with an axonal delay of 0.8ms.
\begin{table}[htbp]
\begin{centering}
\begin{tabular}{|c|c|}
\hline 
Parameter & Value\tabularnewline
\hline 
\hline 
$\vartheta$ & -50mV\tabularnewline
\hline 
$U^{\mathrm{rest}}$ & -60mV\tabularnewline
\hline 
$\tau^{{\rm mem}}$ & 10ms\tabularnewline
\hline 
$\tau^{\mathrm{syn}}$ & 5ms\tabularnewline
\hline 
$\tau^{\mathrm{ref}}$ & 5ms\tabularnewline
\hline 
\end{tabular}
\par\end{centering}

\caption{Neuron model parameters.\label{tab:neuron-model-parameters}}
\end{table}

\subsection{Stimulation paradigms}

Depending on the task at hand, we used two different types of stimuli.
For simulation experiments in which the network had to learn exact
output spike times, we used a set of frozen Poisson spike trains as
input. These stimuli consisted of a single draw of~$n$, where~$n$
is the number of input units, Poisson spike trains of a given duration.
These spike trains were then repeated in a loop and had to be associated
with the target spike train which was consistently aligned to the
repeats of the frozen Poisson inputs. For benchmarking and comparison
reasons, the stimulus and target spike trains shown in this paper
are publicly available as part of the Supervised Spiking Benchmark Suite
(version 71291ea; \citet{zenke_ssbm:_2017}).

For classification experiments we used sets of different stimuli.
Individual stimuli were drawn as random neuronal firing time offsets
from a common stimulus onset time. Stimulus order was chosen randomly
and with randomly varying inter-stimulus-intervals.

\subsection{Plasticity model}

The main ingredients for our supervised learning rule for spiking
neurons (SuperSpike) are summarized in Equation~\ref{eq:learning_rule}
describing the synaptic weight changes. As also eluded to above, 
the learning rule can be interpreted as a nonlinear Hebbian
three factor rule. The nonlinear Hebbian term detects coincidences
between presynaptic activity and postsynaptic depolarization. These
spatiotemporal coincidences at the single synapse $w_{ij}$ are then
stored transiently by the temporal convolution with the causal kernel~$\alpha$.
This step can be interpreted as a synaptic eligibility
trace, which in neurobiology could for instance be implemented as
a calcium transient or a related signaling cascade (cf.\
Fig.~\ref{fig:hidden_layer_concepts}b; \citet{gutig_tempotron:_2006}).
Importantly, the algorithm is causal in the sense that all necessary quantities
are computed online without the need to propagate error signals backwards
through time. Due to this fact, SuperSpike can be interpreted as an
implementation of \gls{rtrl} \citep{williams_learning_1989} for spiking
neural networks.
In the model, all the complexity of neural feedback of learning is absorbed into
the per-neuron signal $e_{i}(t)$. Because it is unclear if and how such error
feedback is signaled to individual neurons in biology here we explored different
strategies which are explained in more detail below. For practical reasons, we
integrate Eq.~\ref{eq:learning_rule} over finite temporal intervals before
updating the weights. The full learning rule can be written as follows:
\begin{equation}
\Delta w_{ij}^k=r_{ij}\,\int_{t_{k}}^{t_{k+1}}\underbrace{e_{i}(s)}_{\text{Error signal}}\,\alpha\ast\left(\underbrace{\sigma^{\prime}(U_{i}(s))}_{\text{Post}}\,\underbrace{\left(\epsilon\ast S_{j}\right)(s)}_{\text{Pre}}\right)\,\,ds\label{eq:learning_rule2}
\end{equation}

In addition to the neuronal dynamics as described in the previous
section, the evaluation of Eq.~\ref{eq:learning_rule} can thus coarsely
be grouped as follows: i) evaluation of presynaptic traces, ii) evaluation
of Hebbian coincidence and computation of synaptic eligibility traces,
iii) computation and propagation of error signals, and iv) integration
of Eq.~\ref{eq:learning_rule} and weight update. We will describe
each part in more detail in the following.

\subsubsection{Presynaptic traces}
Because $\epsilon$ is a double exponential filter, the temporal convolution
in the expression of the presynaptic traces (Eq.~\ref{eq:learning_rule2}),
can be evaluated efficiently online by exponential filtering twice.
Specifically, we explicitly integrate the single exponential trace 
\[
\frac{dz_{j}}{dt}=-\frac{z_{j}}{\tau_{\mathrm{rise}}}+S_{j}(t)
\]
in every time step which is then fed into a second exponential filter
array
\[
\tau_{\mathrm{decay}}\frac{d\tilde{z}_{j}}{dt}=-\tilde{z}_{j}+z_{j}
\]
with $\tilde{z}_{j}(t)\equiv\left(\epsilon\ast S_{j}\right)(t)$ which
now implements the effective shape of a \gls{psp} in the model. In all
cases we chose the time constants $\tau_{\mathrm{rise}}=5\mathrm{ms}$
and $\tau_{\mathrm{decay}}=10\mathrm{ms}$.

\subsubsection{Hebbian coincidence detection and synaptic eligibility traces}
To evaluate the Hebbian term we evaluate the surrogate partial derivative
$\sigma^{\prime}(U_{i})$ in every time step. For efficiency reasons
we use the partial derivative of the negative half of a fast sigmoid 
$f(x)=\frac{x}{1+\left|x\right|}$
which does not require the costly evaluation of exponential functions in every. 
Specifically, we compute
$\sigma^{\prime}(U_{i})=\left(1+\left|h_{i}\right|\right)^{-2}$ with
$h_i\equiv\beta\left(U_{i}-\vartheta\right)$ where $\vartheta$ is
the neuronal firing threshold and $\beta=1\mathrm{mV}$ unless mentioned
otherwise.

We compute the outer product between the delayed presynaptic traces
$\tilde{z}_{j}(t-\Delta)$ and the surrogate partial derivatives
$\sigma^{\prime}(U_{i})(t-\Delta)$ in every time step.
Here the delay $\Delta$ is chosen such that it offsets the 0.8ms axonal
delay which spikes acquire during forward propagation.
Because the presynaptic traces decay to zero quickly 
in the absence of spikes, we approximate them to be exactly zero when
their numerical value drops below machine precision of $10^{-7}$.
This allows us to speed up the computation of the outer product by
skipping these presynaptic indices in the computation. 

To implement the synaptic eligibility trace as given by the temporal
filter~$\alpha$, we filter the values of Hebbian product term with
two exponential filters just like in the case of the presynaptic traces
$z_{j}$ above. It is important to note, however, that these traces
now need to be computed for each synapse $w_{ij}$ which makes the
algorithm scale as $O(n^{2})$ for $n$ being the number of neurons.
This makes it the most obvious target for future optimizations of
our algorithm. Biologically, this complexity could be implemented
naturally simply because synaptic spines are electrical and ionic
compartments in which a concentration transient of calcium or other
messengers decays on short timescales.
For SuperSpike to function properly, it is important that these transients are long
enough to temporally overlap with any causally related error signal $e_{i}(t)$. 
Formally the duration of the transient in the model is
given by the filter kernel shape used to compute the van Rossum distance.
We used a double-exponentially filtered kernel which
has the same shape as a \gls{psp} in the model, but other kernels 
are possible.

\subsubsection{Error signals}
We distinguish two types of error signals: Output error signals and
feedback signals (see below). Output error signals are directly tied to
output units for which a certain target signal exists. Their details
depend on the underlying cost function we are trying to optimize.
Feedback signals, on the other hand, are derived from output error
signals by sending them back to the hidden units. In this study we
used two slightly different classes of output error signals and three different
types of feedback.

At the level of output errors we distinguish between the cases in
which our aim was to learn precisely timed output spikes. In these
cases the output error signals were exactly given by $e_{i}=\alpha\ast(\hat{S}_{i}-S_{i})$
for an output unit $i$. Unless stated otherwise we chose
$\alpha\propto\epsilon$, but normalized to unity.
As can be seen from this expression, the error signal $e_{i}$ only vanishes if
the target and the output spike train exactly match with the temporal precision
of our simulation.
All cost function values were computed online as the root mean square from a
moving average with 10s time constant.

In simulations in which we wanted to classify input spike patterns rather than
generate precisely timed output patterns, we introduced some slack into the
computation of the error signal. For instance, as illustrated  in
Figure~\ref{fig:xor}, we gave instantaneous negative error feedback as described
by $e_{i}=-\alpha\ast S_{i}^{\mathrm{err}}$ for each erroneous additional spike
$S_{i}^{\mathrm{err}}$. However, since for this task we did not want the network
to learn precisely timed output spikes, we only gave a positive feedback signal
$e_{i}=\alpha\ast S_{i}^{\mathrm{miss}}$ at the end of a miss trial, i.e.~when a
stimulus failed to evoke an output spike during the window of opportunity when
it should have (see section on Stimuli above).

\subsubsection{Feedback signals}
We investigated different credit assignment strategies for hidden units.
To that end, hidden layer units received one out
of three types of feedback (cf.\ Fig.~\ref{fig:hidden_layer_concepts}b).
We distinguish between symmetric, random and uniform feedback.
Symmetric feedback signals where computed in analogy to \gls{backprop} as the weighted sum $e_{i}=\sum_k
w_{ki}e_{k}$ of the downstream error signals using the actual feed-forward
weights $w_{ik}$. Note that in contrast to \gls{backprop} the non-local information
of downstream activation functions does not appear in this expression, which is 
closely related to the notion of straight-through estimators \citep{hinton_neural_2012, 
bengio_estimating_2013, baldi_learning_2016}.
Motivated by recent results on feedback alignment \citep{lillicrap_random_2016}, 
random feedback signals were computed as the
random projection $e_{i}=\sum_k b_{ki}e_{k}$ with random coefficients~$b_{ki}$
drawn from a normal distribution with zero mean and unit variance. 
This configuration could be implemented, for instance, by individual neurons
sensing differential neuromodulator release from a heterogeneous population of modulatory
neurons.
Finally, in the case of uniform feedback all weighting coefficients were simply
set to one $e_{i}=\sum_k e_{k}$ corresponding closest to a single global third
factor distributed to all neurons, akin to a diffuse neuromodulatory signal.

\subsubsection{Weight updates}
To update the weights, the time continuous time series corresponding
to the product of error/feedback signal and the synaptic eligibility
traces $\lambda_{ij}$ were not directly added to the synaptic weights,
but first integrated in a separate variable $m_{ij}$
in chunks of $t_{b}=0.5257\mathrm{s}$. Specifically, we computed
$m_{ij}\rightarrow m_{ij}+g_{ij}$ with $g_{ij}(t)=e_{i}(t)\,\lambda_{ij}(t)$
at each time step. For stimuli exceeding the duration $t_{b}$ this
can thus be seen as the continuous time analogue to mini batch optimization.
We chose $t_{b}$ on the order of half a second as a good compromise
between computational cost and performance for synaptic updates and
added 257 simulation time steps to minimize periodic alignment of
the update step with the stimulus. At the end of each interval $t_{b}$,
all weights were updated according to $w_{ij}\rightarrow w_{ij}+r_{ij}m_{ij}$
with the per parameter learning rate $r_{ij}$. In addition to that,
we enforced the constraint for individual weights to remain in the
interval $-0.1<w_{ij}<0.1$. After updating the weights, the variables
$m_{ij}$ were reset to zero.

\subsubsection{Per-parameter learning rates}
To facilitate finding the right learning rate and the speed up training
times in our simulations, we implement a per-parameter learning rate
heuristic. To compute the per-parameter learning rate, in addition
to $m_{ij}$ we integrated another auxiliary quantity $v_{ij}\rightarrow\max(\gamma v_{ij},g_{ij}^{2}$).
Here $\gamma=\exp(-\nicefrac{\Delta}{\tau_{\mathrm{rms}}})$
ensures a slow decay of $v_{ij}$ for $g_{ij}=0$. Consequently,
$v_{ij}$ represents an upper estimate of the variance of
the surrogate gradient for each parameter on the characteristic timescale
$\tau_{\mathrm{rms}}$. With these definitions, the per-parameter learning rate was defined
as $r_{ij}\equiv\frac{r_{0}}{\sqrt{v_{ij}}}$. This choice is motivated
by the RMSprop optimizer which is commonly used in the field of deep
learning \citep{hinton_neural_2012}. However, RMSprop computes a moving exponential average over
the $g_{ij}^{2}$. We found that introducing the max function rendered
training more stable while simultaneously yielding excellent convergence
times. We call this slightly modified version ``RMaxProp''
(compare also AdaMax \citep{kingma_adam:_2014}).
Finally, the parameter~$r_{0}$ was determined via grid search over the values
$(10, 5, 1, 0.5, 0.1)\times10^{-3}$.

\subsubsection{Regularization term}
In some experiments with random feedback we added a heterosynaptic
regularization term to the learning rule of the hidden layer weights
to avoid pathologically high firing rates. In these experiments the
full learning rule was 

\begin{equation}
\frac{\partial w_{ij}^{\mathrm{hid}}}{\partial t}=r_{ij}\,\int_{t_{k}}^{t_{k+1}}\underbrace{e_{i}(s)}_{\text{Error signal}}\,\left(\alpha\ast\left(\underbrace{\sigma^{\prime}(U_{i}(s))}_{\text{Post}}\,\underbrace{\left(\epsilon\ast S_{j}\right)(s)}_{\text{Pre}}\right)\,-\underbrace{\rho w_{ij}e_{i}(s)z_{i}^{\mathrm{4}}}_{\mathrm{Regularizer}}\right)\,ds\label{eq:learning_rule2-1}
\end{equation}
where we introduced the regularization strength parameter $\rho$.
We made the regularization term explicitly dependent on the square
of the error signal to ensure it would be zero in cases where the
task was solved perfectly. Moreover, we used the fourth power of an
exponential synaptic trace $z_{i}$ which evolved according to the
following differential equation

\[
\frac{dz_{i}}{dt}=-\frac{z_{i}}{\tau_{\mathrm{het}}}+S_{i}(t)
\]
the weight and forth power rate-dependence was motivated from previous
work \citep{zenke_diverse_2015} and to regularize high firing rates more strongly.

\section{Numerical experiments}
\begin{figure}[thbp]
	\centering
	\includegraphics[scale=1.0]{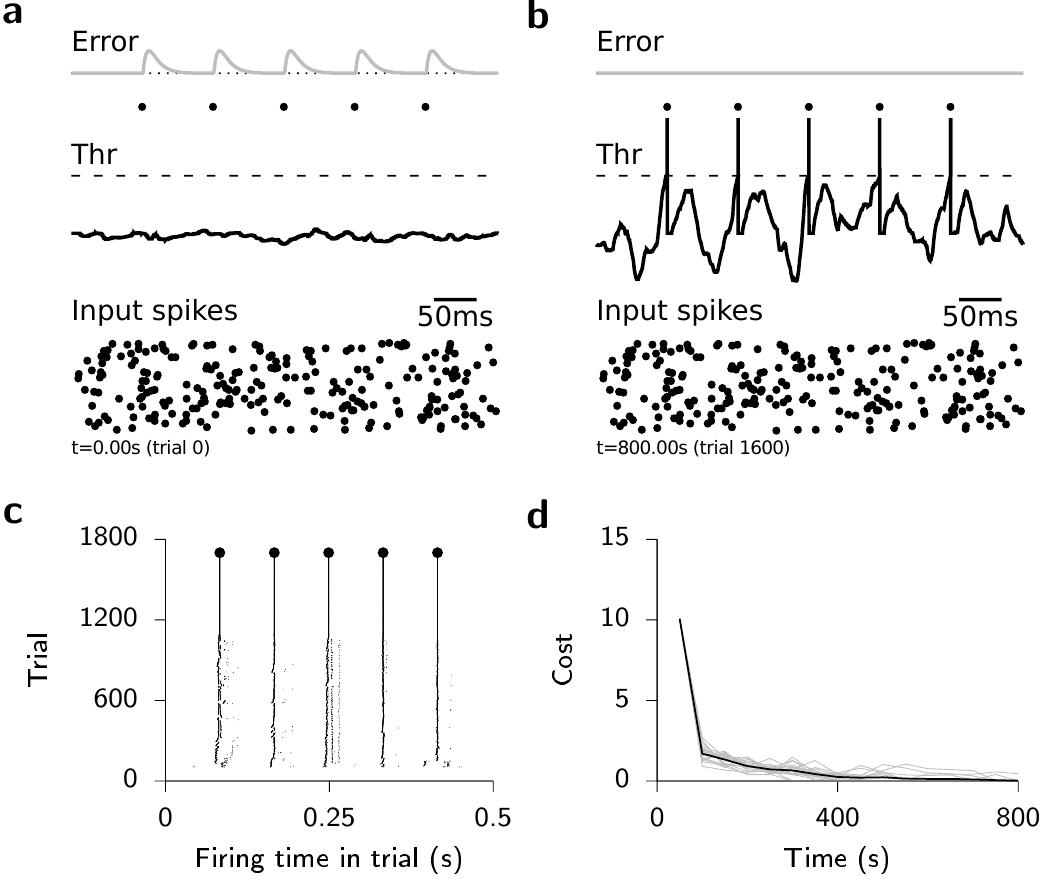}
\caption{SuperSpike learns precisely timed output spikes for a single output neuron. 
	(\textbf{a})~Snapshot of initial network activity. 
	Bottom panel: Spike raster of the input activity.
	Middle panel: The membrane potential of the output neuron (solid
	black line) and its firing threshold (dashed line). Target spikes
	are shown as black points. 
	Top panel: Error signal (gray solid line).
	Zero error is indicated for reference as dotted line. 
	(\textbf{b})~Same as in~(a), but after 800s of SuperSpike learning. 
	(\textbf{c})~Spike timing plot showing the temporal evolution of per-trial firing times
	(\textbf{d})~Learning curves of 20 trials (gray) as well as their mean
	(black line) during training. \label{fig:single_unit}}
\end{figure}

To test whether Equation~\ref{eq:learning_rule} could be used to train
a single neuron to emit a predefined target spike pattern, we simulated
a single \gls{lif} neuron which received a set of 100 spike
trains as inputs. The target spike train was chosen as 5~equidistant
spikes over the interval of 500ms. The inputs were drawn as Poisson
spike trains that repeated every 500ms. We initialized
the weights in a regime were the output neuron only showed sub-threshold
dynamics, but did not spike (Fig.~\ref{fig:single_unit}a). 
Previous methods, starting from this quiescent state, 
would require the introduction of noise to generate spiking, 
which would in turn retard the speed with which precise output spike times could be learned. 
Finally, weight updates were computed by evaluating the integral in Eq.~\ref{eq:learning_rule}
over a fixed interval and scaling the resulting value with the learning
rate (Methods). After 500 trials, corresponding to 250s of simulated
time, the output neuron had learned to produce the desired output
spike train (Fig.~\ref{fig:single_unit}b). However, fewer trials could  
generate good approximations to
the target spike train (Fig.~\ref{fig:single_unit}c).

\subsection{Learning in multi-layer spiking neural networks}
\begin{figure}[thbp]
\centering
\includegraphics[scale=1.0]{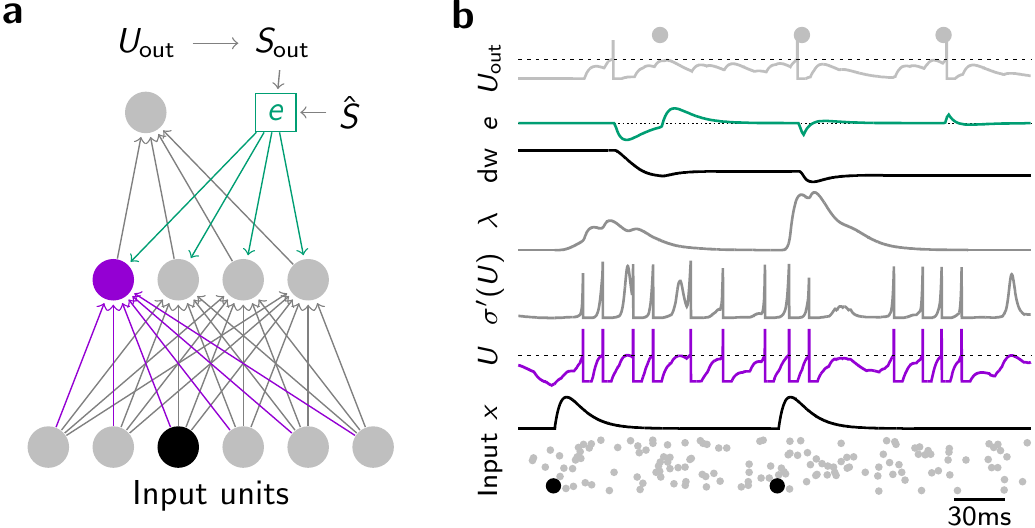}
\caption{(\textbf{a})~Schematic illustration of SuperSpike learning in a network
with a hidden layer. Spikes generated by the lower input layer are
propagated through the hidden layer in the middle to the output layer
at the top. 
(\textbf{b})~Temporal
evolution of the dynamical quantities involved in updating a single synaptic
weight from an input to a hidden layer unit. For brevity we have suppressed the neuron indices on all the variables.
Input spikes (bottom panel) and their associated post-synaptic potentials $x$ sum
to the membrane voltage in the hidden unit (purple). Further
downstream, the spikes generated in the hidden layer sum at the output unit 
($U_{\mathrm{out}}$). Finally, the
error signal $e$ (shown in green) is computed from the output spike
train. It modulates learning of the output weights and is propagated 
back to the hidden layer units through feedback weights. Note that the
error signal $e$ is strictly causal. The product of presynaptic activity
($x$) with the nonlinear function
$\sigma^{\prime}(U)$ is further filtered in time by $\alpha$ giving 
rise to the synaptic eligibility trace $\lambda$. 
In a biological scenario  $\lambda$ could for instance be manifested
as a calcium transient at the synaptic spine. Finally,
temporal coincidence between $\lambda$ and the error signal $e$
determines the sign and magnitude of the plastic weight changes $dw$.
\label{fig:hidden_layer_concepts}}
\end{figure}

Having established that our rule can efficiently
transform complex spatiotemporal input spike patterns to precisely timed output spike trains in a network without hidden units, 
we next investigated how well the same rule would perform in multilayer networks.
The form of Equation~\ref{eq:learning_rule} suggests a straight
forward extension to hidden layers in analogy to \gls{backprop}.
Namely, we can use the same learning rule
(Eq.~\ref{eq:learning_rule}) for hidden units, with the modification that
that $e_{i}(t)$ becomes a complicated function which depends on the 
weights and future activity of all downstream neurons. However, this 
non-locality in space and time presents serious problems, both in terms of biological plausibility and technical feasibility. Technically, this computation requires either backpropagation through time through the \gls{psp}
kernel or the computation of all relevant quantities online as in the case of
\gls{rtrl}.
Here we explore the latter approach since
our specific choice of temporal kernels allows us to compute all relevant dynamic quantities  
and error signals online (Fig.~\ref{fig:hidden_layer_concepts}b). 
In our approach, error signals are distributed directly through a feedback matrix to the hidden
layer units (Fig.~\ref{fig:hidden_layer_concepts}a). 
Specifically, this means that the output error signals are neither  
propagated through the actual nor the ``soft'' spiking nonlinearity. This
idea is closely related to the notion of straight-through estimators in machine learning
\citep{hinton_neural_2012, bengio_estimating_2013, baldi_learning_2016}.
We investigated different configurations of the feedback matrix, which can be 
either (i)~symmetric (i.e.\ the transpose of the feedforward weights), as in the
case of \gls{backprop}, (ii)~random as motivated by the recent results on feedback alignment \citep{lillicrap_random_2016} or (iii)~uniform, corresponding closest to a single global third factor distributed to all neurons, akin to a diffuse neuromodulatory signal.  

\begin{figure}[thbp]
\centering
	\includegraphics[scale=1.0]{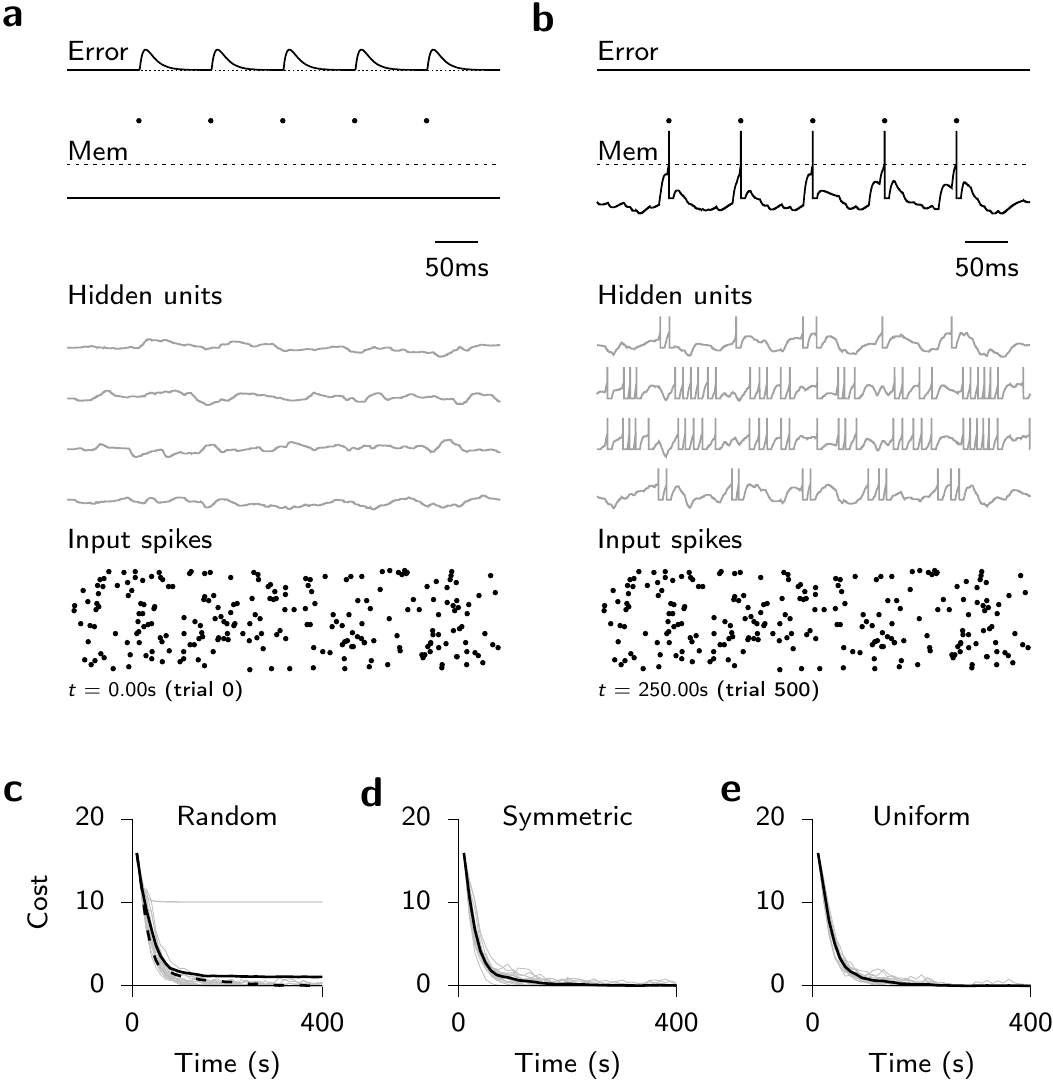}
	\caption{SuperSpike learning with different types of feedback allows to
	train multi-layer networks. 
	(\textbf{a})~Network activity at the initial trial at reference time
	$t=0\mathrm{s}$. The bottom panel shows the membrane potential traces of the
	four hidden units. The membrane potential of the output unit is shown in the
	middle. The dashed line is the output neuron firing threshold. The points
	correspond to target firing times and the top plot shows the error signal at
	the output layer.  Hidden units receive the same input spikes as shown in
	Fig.~\ref{fig:single_unit}a.
	(\textbf{b})~Same as (a), but after 250s of training. The two hidden units
	which have started to respond to the repeating input spiking pattern are the
	ones with positive feedback weights, whereas the two hidden neurons which
	receive negative feedback connections from the output layer (middle traces)
	respond mostly at the offset of the repeating stimulus. 
	(\textbf{c})~Learning curves of networks 
	trained with random feedback connections. 
	Gray lines correspond to single trials and the black line to the average. 
	The dashed line is the same average but for a network with 8~hidden layer units.
	(\textbf{d})~Same as (d), but for a network with symmetric feedback connections. 
	(\textbf{e})~Same as (d--e), but for uniform ``all one'' feedback
	connections.
\label{fig:single_hidden}}
\end{figure}

We first sought to replicate the task shown in Figure~\ref{fig:single_unit},
but with the addition of a hidden layer composed of 4~\gls{lif} neurons.
Initially, we tested learning with random feedback. 
To that end, feedback weights were drawn from a zero mean unit variance Gaussian 
and their value remained fixed during the entire simulation.
The synaptic feedforward weights were also initialized randomly
at a level at which neither the hidden units nor the output unit fired
a single spike in response to the same input spike trains as used
before (Fig.~\ref{fig:single_hidden}a). After training the network
for 40s, some of the hidden units had started to fire spikes in response
to the input. Similarly, the output neuron had started to fire at
intermittent intervals closely resembling the target spike train (not shown).
Continued training on the same task for a total of 250s lead to a
further refinement of the output spike train and more differentiated
firing patterns in a subset of the hidden units (Fig.~\ref{fig:single_hidden}b). 

Although, we did not restrict synaptic connectivity to obey Dale's
principle, in the present example with random feedback all hidden neurons with positive feedback
connections ended up being excitatory, whereas neurons with negative
feedback weights generally turned out to be inhibitory at the end
of training. These dynamics are a direct manifestation of ``feedback
alignment'' aspect of random feedback learning \citep{lillicrap_random_2016}. 
Because the example shown in Figure~\ref{fig:single_hidden} does not strictly
require inhibitory neurons in the hidden layer, in many cases the neurons with
negative feedback remained quiescent or at low activity levels at the end of learning
(Fig.~\ref{fig:single_hidden}b--c). 

Learning was successful for different initial conditions, although
the time for convergence to zero cost varied (Fig.~\ref{fig:single_hidden}d).
We did encounter, however, a few cases in which the network completely
failed to solve the task. These were the cases in which \emph{all}
feedback connections happened to be initialized with a negative value
(Fig.~\ref{fig:single_hidden}c). This eventuality could be made
very unlikely, however, by increasing in the number of hidden units
(Fig.~\ref{fig:single_hidden}c). Other than that, we did not find
any striking differences in performance when we replaced the random
feedback connections by symmetric (Fig.~\ref{fig:single_hidden}d)
or uniform ``all one'' feedback weights (Fig.~\ref{fig:single_hidden}e).

\begin{figure}[thbp]
	\centering{}
	\includegraphics[scale=1.0]{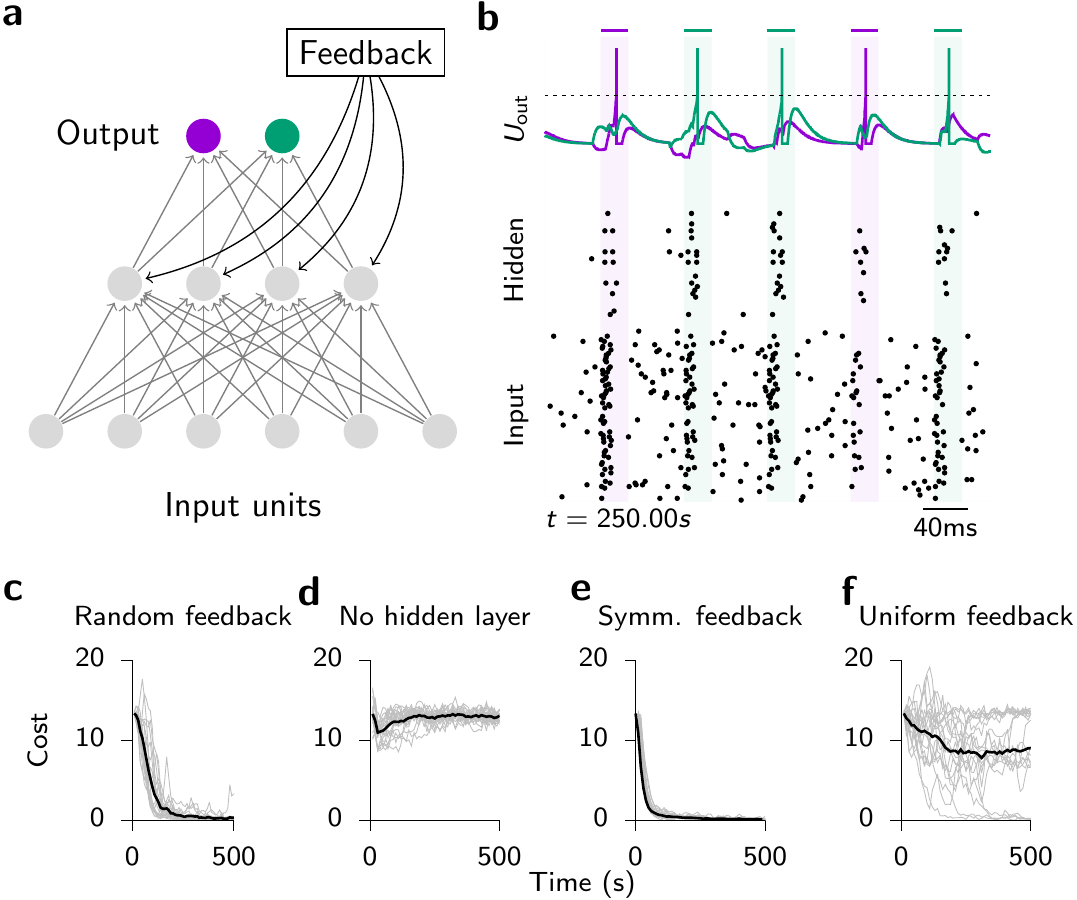}
	\caption{Network trained to solve a non-linearly separable classification problem with
	noisy input neurons. 
	(\textbf{a})~Sketch of network layout with two output units and four hidden
	units. 
	(\textbf{b})~Snapshot of network activity at the end of training. Four input
	patterns from two non-linearly separable classes are presented in random
	order (gray shaded areas). In between stimulus periods input neurons spike
	randomly with 4Hz background firing rate.
	(\textbf{c})~Learning curves of 20 trials with different random
	initializations (gray) for a network without hidden layer which cannot solve
	the task. The average of all trials is given by the black line. 
	(\textbf{d})~Same as (d), but for a network without hidden units which
	receive random feedback during training. 
	(\textbf{e})~Same as (d), but for symmetric feedback. 
	(\textbf{f})~Same as (d), but for uniform (``all ones'') feedback
	connections. \label{fig:xor}}
\end{figure}
The previous task was simple enough such that solving it did not require a
hidden layer. We therefore investigated whether SuperSpike could also learn
to solve tasks that cannot be solved by a network without hidden units. To that end, we
constructed a spiking \textit{exclusive-or} task in which four different spiking 
input patterns had to be separated into two classes.
In this example we used 100~input units although the effective dimension of the problem 
was two by construction.
Specifically, we picked three non-overlapping sets of input neurons with
associated fixed random firing times in a 10ms window.
One set was part of all patterns and served as a time reference. 
The other two sets were combined to yield the four input patterns of 
the problem. Moreover, we added a second readout neuron each corresponding to one
of the respective target classes (Fig.~\ref{fig:xor}a).
The input patterns were given in random order as short bouts of spiking
activity at random inter-trial-intervals during which input neurons were firing
stochastically at 5Hz (Fig.~\ref{fig:xor}b).
To allow for the finite propagation time through the network, we relaxed
the requirement for precise temporal spiking and instead required
output neurons to spike within a narrow window of opportunity which was 
was aligned with and outlasted each stimulus by 10ms. 
The output error signal was zero unless the correct output neuron
failed to fire within the window. In this case an error signal
corresponding to the correct output was elicited at the end of the window. 
At any time an incorrect spike triggered immediate negative feedback. 
We trained the network comparing different types of feedback. 
A network with random feedback quickly learned
to solve this task with perfect accuracy (Fig.~\ref{fig:xor}b--c),
whereas a network without hidden units was unable to solve the task
(Fig.~\ref{fig:xor}d). Perhaps not surprisingly, networks with symmetric
feedback connections also learned the task quickly and overall their
learning curves were more stereotyped and less noisy (Fig.~\ref{fig:xor}e),
whereas networks with uniform feedback performed worse
on average (Fig.~\ref{fig:xor}f). 
Overall these results illustrate
that temporally coding spiking multi-layer networks can be trained
to solve tasks which cannot be solved by networks without hidden layers.
Moreover, these results show that random feedback is
beneficial over uniform feedback in some cases.

\subsection{Limits of learning with random feedback}
\begin{figure}[thbp]
	\centering{}
	\includegraphics[scale=1.0]{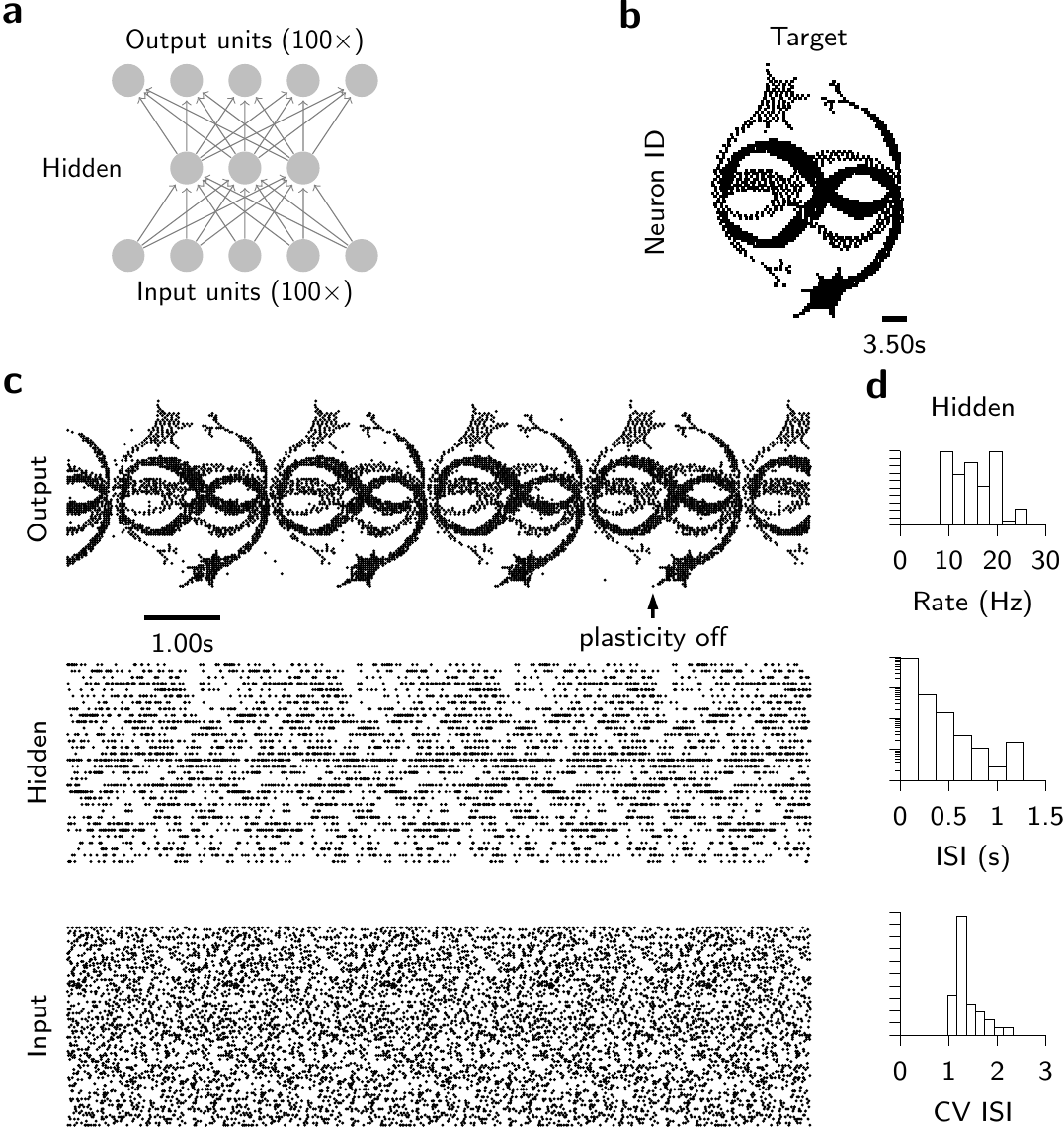}
	\caption{Learning of complex spatiotemporal spike pattern transformations. 
	(\textbf{a})~Spike raster of target firing pattern of 100~output neurons.
	The whole firing pattern has a duration of 3.5s. (\textbf{b})~Schematic
	illustration of the network architecture.
	(\textbf{b})~Spike raster of target firing pattern for reference.
	(\textbf{c})~Snapshot of network activity of the network with symmetric
	feedback after 1000s of SuperSpike learning. Bottom panel: Spike raster of
	repeating frozen Poisson input spikes. Middle panel: Spike raster of hidden
	unit spiking activity. Top panel: Spike raster of output spiking activity.
	The black arrow denotes the point in time at which SuperSpike learning is
	switched off which freezes the spiking activity of the fully deterministic
	network. 
	(\textbf{d})~Histograms of different firing statistics of hidden layer
	activity at then of learning. Top: Distribution of firing rates. Middle:
	Inter-spike-interval (ISI) distribution on semi-log axes. Bottom:
	Distribution of coefficient of variation (CV) of the ISI distribution. 
	\label{fig:auryn}}
\end{figure}

\begin{figure}[thbp]
	\centering{}
	\includegraphics[scale=1.0]{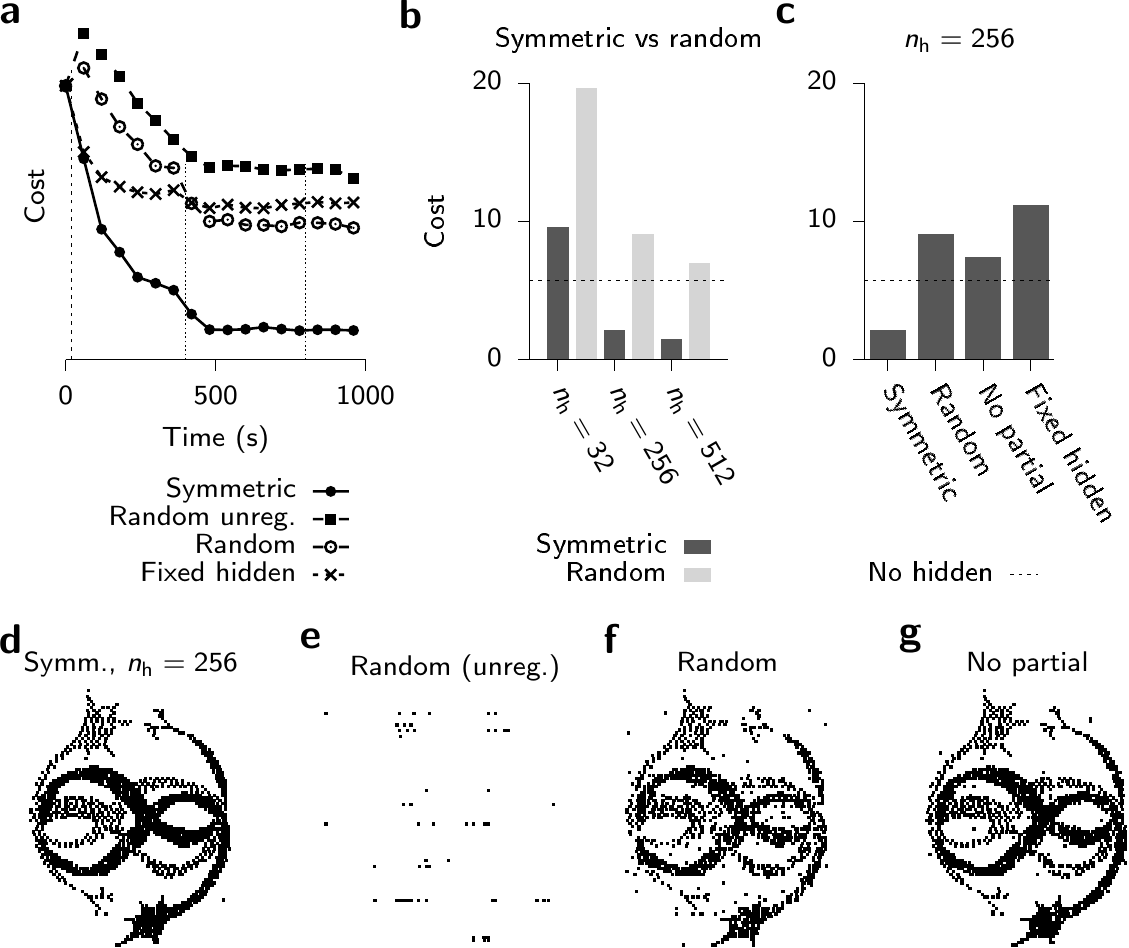}
	\caption{Learning of spatiotemporal spike patterns. 
	(\textbf{a})~Learning curves for networks with symmetric feedback or no
	hidden layer respectively. 
	Learning was activated at the dashed line, whereas the learning rate was reduced by a factor 
	of 10 at each dotted line.
	(\textbf{b})~Minimum cost after convergence for different feedback
	strategies and varying numbers of hidden units.
	Dashed line: Performance of a network without hidden units. 
	(\textbf{c})~Minimum cost after convergence for symmetric feedback
	with a fixed number of $n_\mathrm{h}=256$.
	Dashed line: Performance of a network without hidden units. 
	No partial: Learning rule, but without voltage nonlinearity.
	Fixed hidden: Learning in hidden units disabled.
	(\textbf{d})~Spike raster snapshots of output activity after learning for 
	symmetric feedback ($n_\mathrm{h}=256$). 
	(\textbf{e})~Like (d), but for unregularized random feedback.
	(\textbf{f})~Like (e), but with additional heterosynaptic regularization (Methods).
	(\textbf{g})~Like (d), but without voltage nonlinearity.
	\label{fig:auryn2}}
\end{figure}

All tasks considered so far were simple enough that they could be solved by
most three layer networks with zero error for all types of feedback signals.
We hypothesized that the observed indifference to the type of feedback could
be due to the task being too simple. 
To test whether this picture would change for a more challenging
task we studied a network with 100~output neurons which had to learn a
3.5 second-long complex spatiotemporal output pattern from cyclically repeating 
frozen Poisson noise (Methods). 
Specifically, we trained a three layer \gls{snn} with 100~input, 100~output, and
different numbers of hidden neurons (Fig.~\ref{fig:auryn}a). 
Within 1000s of training with symmetric feedback connections, a network with
32~or more hidden units could learn to emit an output spike pattern which
visually matched to the target firing pattern
(Fig.~\ref{fig:auryn}b,c). 
After successful learning, hidden unit activity was irregular and at
intermediate firing rates of 10--20Hz with a close to exponential
inter-spike-interval distribution (Fig.~\ref{fig:auryn}d).
However, the target pattern was not learned perfectly as evidenced by
a number of spurious spikes (Fig.~\ref{fig:auryn}b) and 
a non-vanishing van Rossum cost (Fig.~\ref{fig:auryn2}a).

On the same task, a simulation with random feedback yielded
substantially worse 
performance (Fig.~\ref{fig:auryn2}a) and the output pattern became close to
impossible to recognize visually (Fig.~\ref{fig:auryn2}e).
As expected, results from uniform feedback were worst (not shown), and hence this
option will not be considered in the following.
Notably, the random feedback case performs worse than a network which was trained
without a hidden layer (Fig.~\ref{fig:auryn2}b).
Since, we observed abnormally high firing rates in hidden layer neurons in
networks trained with random feedback, we 
wondered whether performance could be improved through the addition of
a heterosynaptic weight decay (Methods) which acts as an activity regularizer \citep{zenke_diverse_2015}.
The addition of such a heterosynaptic weight decay term notably
improved learning performance (Fig.~\ref{fig:auryn2}a)
and  
increased the visual similarity of the output patterns
(Fig.~\ref{fig:auryn2}d--f).
However, even this modified learning rule did not achieve comparable performance
levels to a symmetric-feedback network.
Importantly, for the hidden layer sizes we tested, random feedback networks did
not even achieve the same performance levels as networks without a hidden layer, whereas
symmetric feedback networks did (Fig.~\ref{fig:auryn2}b).
Not surprisingly, networks with wider hidden layers performed superior to
networks with fewer hidden units, but networks with random feedback performed
consistently worse than their counterparts trained with symmetric feedback
(Fig.~\ref{fig:auryn2}b). 
Finally, when we trained the network using symmetric feedback with a learning
rule in which we disabled the nonlinear voltage dependence by setting the
corresponding term to~1, the output pattern was degraded 
(``no partial'' in Fig.~\ref{fig:auryn2}g; cf.\ Eq.~\ref{eq:learning_rule}).

These results seem to confirm our intuition, that for more challenging tasks the
nonlinearity of the learning rule, firing rate regularization, and non-random
feedback seem to become more important to achieving good performance on the type
of spatiotemporal spike pattern transformation tasks we considered here.

\section{Discussion}
In this paper we have derived a 
three factor learning rule to train
deterministic multi-layer \glspl{snn} of \gls{lif} neurons. 
Moreover, we have assessed the impact of different types of feedback credit assignment strategies for the hidden units, notably symmetric, random, and uniform.
In contrast to previous work \citep{pfister_optimal_2006, fremaux_functional_2010, gardner_learning_2015}, 
we have used a deterministic surrogate gradient approach instead of
the commonly used stochastic gradient approximations.
By combining this rule with ideas of straight-through estimators \citep{hinton_neural_2012, bengio_estimating_2013}
and feedback alignment \citep{lillicrap_random_2016, baldi_learning_2016}, we could 
efficiently train and study precisely timed spiking dynamics in multi-layer networks of
deterministic \gls{lif} neurons without relying on the introduction of extraneous and unavoidable noise present in stochastic models, noise which generally impedes the ability to learn precise spatiotemporal spike-pattern transformations. 

The weight update equation of SuperSpike constitutes a 
voltage-based nonlinear Hebbian three factor rule with individual 
synaptic eligibility traces. 
These aspects each have direct biological interpretations.
For instance, a nonlinear voltage dependence has been reported ubiquitously by numerous studies on
Hebbian long-term plasticity induction in hippocampus and cortex \citep{artola_different_1990, feldman_spike-timing_2012}.
Also, the window of temporal coincidence detection in our model is in good 
agreement with that of \gls{stdp} \citep{feldman_spike-timing_2012}.
Moreover, the time course of the eligibility traces could be interpreted as a local calcium transient at the synaptic spine level.
Finally, the multiplicative coupling of the error signal with the eligibility
trace could arise from neuromodulators \citep{izhikevich_solving_2007,
pawlak_timing_2010, fremaux_neuromodulated_2016, kusmierz_learning_2017}.
However, instead of only one global feedback signal, our work highlights the necessity 
of a higher dimensional neuromodulatory or electrical feedback signal for learning 
potentially with some knowledge of the feedforward pathway. 
The biological exploration of such intelligent neuromodulation, as well as extensions of our 
approach to deeper and recurrent \glspl{snn}, are left as intriguing directions for future work.

\section*{Acknowledgements}
The authors would like to thank Subhy Lahiri and Ben Poole for helpful
discussions.
FZ was supported by the SNSF (Swiss National Science Foundation) and the Wellcome Trust.
SG was supported by the Burroughs Wellcome, Sloan, McKnight, Simons and James S. McDonnell foundations and the Office of Naval Research.

\bibliographystyle{neco}
\bibliography{library}

\end{document}